\documentclass[onecolumn]{webofc}
\usepackage[varg]{txfonts}   
\usepackage{hyperref}
\usepackage{color}
\usepackage{enumitem}
\usepackage{framed}
\usepackage{microtype}
\usepackage{xcolor}
\usepackage{booktabs}
\usepackage{subcaption}
\usepackage{setspace}
\usepackage[
  shortcuts,%
  acronym,%
  nonumberlist,%
  nogroupskip,%
  nopostdot]{glossaries}
\usepackage[
    detect-none,%
    binary-units]{siunitx}
\usepackage{orcidlink}

\usepackage[textsize=tiny]{todonotes}
\definecolor{CiteColor}{rgb}{0, 0, 0.55}
\definecolor{LinkColor}{rgb}{0.2, 0.2, 0.2}
\definecolor{URLColor}{rgb}{0.62745098, 0.1254902 , 0.94117647}
\hypersetup{
  linkcolor=LinkColor,
  citecolor=CiteColor,
  urlcolor=URLColor,
  colorlinks=true
}

\setacronymstyle{long-short}

\newacronym{atlas}{ATLAS}{A Toroidal LHC apparatus}
\newacronym{api}{API}{Application Programming Interface}
\newacronym{doe}{DOE}{U.S. Department of Energy}
\newacronym{cmssw}{CMSSW}{CMS software}
\newacronym{cms}{CMS}{Compact Muon Solenoid}
\newacronym{em}{EM}{electromagnetic}
\newacronym{gpu}{GPU}{}
\newacronym{hep}{HEP}{high energy physics}
\newacronym{hllhc}{HL-LHC}{High-Luminosity Large Hadron Collider}
\newacronym{hgcal}{HGCAL}{High Granularity Calorimeter}
\newacronym{hpc}{HPC}{high performance computing}
\newacronym{lhc}{LHC}{Large Hadron Collider}
\newacronym{mc}{MC}{Monte Carlo}
\newacronym{msc}{MSC}{multiple scattering}
\newacronym{olcf}{OLCF}{Oak Ridge Leadership Computing Facility}
\newacronym{ornl}{ORNL}{Oak Ridge National Laboratory}
\newacronym{orange}{ORANGE}{Oak Ridge Adaptable Nested Geometry Engine}
\newacronym{sd}{SD}{sensitive detector}
\newacronym{sdk}{SDK}{software development kit}
\newacronym{prng}{PRNG}{pseudo-random number generator}
\newacronym{mt}{MT}{multithreaded}
\newacronym{mimd}{MIMD}{multiple instruction, multiple data}

\begin{document}

\title{Accelerating detector simulations with Celeritas: profiling and performance optimizations%
  %
}

\author{%
  \firstname{Amanda L.} \lastname{Lund}\inst{1}\orcidlink{0000-0002-8316-0709}\fnsep\thanks{\email{alund@anl.gov}}
  \and
  \firstname{Julien} \lastname{Esseiva}\inst{2}\orcidlink{0009-0002-1119-4851}
  \and
  \firstname{Seth R.} \lastname{Johnson}\inst{3}\orcidlink{0000-0003-1504-8966}
  \and
  \firstname{Elliott} \lastname{Biondo}\inst{3}\orcidlink{0000-0002-9088-1360}
  \and
  \firstname{Philippe} \lastname{Canal}\inst{4}\orcidlink{0000-0002-7748-7887}
  \and
  \firstname{Thomas} \lastname{Evans}\inst{3}\orcidlink{0000-0001-5743-3788}
  \and
  \firstname{Hayden} \lastname{Hollenbeck}\inst{6}\orcidlink{0000-0000-0000-0000}
  \and
  \firstname{Soon Yung} \lastname{Jun}\inst{4}\orcidlink{0000-0003-3370-6109}
  \and
  \firstname{Guilherme} \lastname{Lima}\inst{4}\orcidlink{0000-0003-4585-0546}
  \and
  \firstname{Ben} \lastname{Morgan}\inst{5}\orcidlink{0000-0003-3604-0883}
  \and
  \firstname{Stefano C.} \lastname{Tognini}\inst{3}\orcidlink{0000-0001-9741-6608}
}%
\institute{%
  Argonne National Laboratory, Lemont, IL, USA
  \and
  Lawrence Berkeley National Laboratory, Berkeley, CA, USA
  \and
  Oak Ridge National Laboratory, Oak Ridge, TN, USA
  \and
  Fermi National Accelerator Laboratory, Batavia, IL, USA
  \and
  University of Warwick, Coventry, United Kingdom
  \and
  University of Virginia, Charlottesville, VA, USA
}%
\abstract{%
Celeritas is a \acs{gpu}-optimized \ac{mc} particle transport code designed to meet the growing computational demands of next-generation \ac{hep} experiments. It provides efficient simulation of \ac{em} physics processes in complex geometries with magnetic fields, detector hit scoring, and seamless integration into Geant4-driven applications to offload \ac{em} physics to \acs{gpu}s. Recent efforts have focused on performance optimizations and expanding profiling capabilities. This paper presents some key advancements, including the integration of the Perfetto system profiling tool for detailed performance analysis and the development of track-sorting methods to improve computational efficiency.
}

\maketitle

\section{Introduction}
\label{sec:introduction}

Detector simulation is an essential component of the design and analysis of \ac{hep} experiments. However, it requires significant computing resources, and with the \ac{hllhc} upgrade these demands will grow beyond the capabilities of conventional computing systems. Celeritas~\cite{celeritas}, a rapidly developing \ac{gpu}-enabled \ac{mc} particle transport code, aims to address these challenges by leveraging high-performance heterogeneous architectures. Over the past several years, Celeritas has implemented the core infrastructure and capabilities~\cite{johnson_novel_2021, tognini_celeritas_2022, johnson_chep_2023, celeritas_rd_report_2024, celeritas_midterm_2024} needed to accelerate detector simulation, including \ac{em} physics processes, full device-based navigation in complex geometries and in the presence of magnetic fields, detector hit scoring, and an \acs{api} for integrating into existing Geant4~\cite{geant4-2003} applications and experimental frameworks. This paper presents recent developments in performance optimization and profiling, including the integration of the Perfetto \cite{perfetto} system profiling and trace analysis tool.

\section{Stepping algorithm}

Adapting methods originally designed for CPU-based codes like Geant4 requires significant algorithmic modifications. One key area of change is the overall parallelization strategy. Traditionally, particles are tracked sequentially from birth, undergoing a series of boundary crossings and interactions until termination. Because individual particle histories are independent, concurrency is achieved by distributing groups of particles to threads. While this performs well on CPUs, it is an inherently \ac{mimd} approach that inhibits data-level parallelism on \ac{gpu}s and can lead to significant thread divergence.

A strategy that has proven successful in \ac{gpu}-enabled neutral particle \ac{mc} codes~\cite{shift_gpu_2019, openmc_gpu_2024}, and the approach adopted by Celeritas, restructures the stepping loop so the fundamental unit of work is a specific particle action rather than an entire particle history. The CPU coordinates an outer loop over a sequence of actions---such as step limit calculation, propagation, and discrete interaction---each executed as a kernel launch on a large array of active tracks. Each iteration of this loop is a \emph{step iteration}. These smaller, specialized kernels use fewer registers, achieve higher occupancy, and reduce thread divergence by ensuring most threads are performing similar tasks.

The active tracks are stored as a struct of arrays, with each array containing per-track state data such as the current energy, position, particle type, and ID of the next action. Instead of partitioning state data into separate particle buffers or action queues, each kernel operates on all tracks, using masking to select particles that participate in a given action.

Secondaries produced in interactions are pushed to a separate buffer of \emph{track initializers}---a lightweight \texttt{struct} containing the minimal data needed to initialize a new track. At the start of each step iteration, vacant slots in the state array are initialized deterministically to ensure reproducibility. \Ac{prng} states are assigned to track slots, meaning each track's random number sequence is dictated by its index in the state array. Because Celeritas executes nearly the same code on the \ac{gpu} and the CPU, reproducibility is maintained across architectures, provided the same number of track slots is used.

\section{Performance optimizations}

Efficient execution on \ac{gpu}s relies on optimizing control flow and memory access patterns: performance is maximized when adjacent threads follow the same execution paths and access contiguous memory. However, the thread masking in kernels and the arbitrary ordering of tracks introduce branching and inefficient memory transactions. To address this, we have explored strategies to reorder tracks, grouping together those undergoing similar actions. Along the same lines, additional optimizations have been investigated to improve load balancing and reduce synchronization overhead.

\subsection{Reindexing track slots}

Because sorting track state data at every step iteration is prohibitively expensive, we explored sorting only the track slot \emph{indices} and accessing the tracks through an indirection. Each thread \texttt{tid} retrieves its track state via \texttt{state[state\_indices[tid]]} using the lookup array \texttt{state\_indices}. Only \texttt{state\_indices} is sorted, ensuring adjacent threads share properties such as particle type or status (i.e., whether a track is active).

This approach aims to reduce branching and kernel grid sizes when grouping tracks based on action or status. The hypothesis is that, given significant thread divergence, memory access is already non-coalesced, and thus reindexing does not substantially degrade the access pattern. In practice, this generally leads to a slowdown compared to unsorted execution, though the performance penalty is smaller for problems with complex geometries and a large number of active tracks.

The best performance is observed when sorting tracks by their along-step action---specifically, distinguishing between neutral particles propagating in a straight line and charged particles undergoing multiple scattering, possibly in a magnetic field. Kernel analysis using NVIDIA Nsight Systems confirms that thread divergence decreases, with up to $38\%$ fewer instructions executed per along-step kernel, depending on the problem geometry. However, this indexing model significantly hinders coalesced memory access, increasing global memory read/write operations up to $344\%$, which negates any potential gains from executing fewer instructions.

\subsection{Partitioning track data}

A second reordering strategy to reduce branching involves sorting track \emph{data} by charged and neutral particles without explicitly sorting the track arrays. This is achieved by partitioning the track initializers scheduled to fill vacant slots in the next step iteration and assigning the charged and neutral initializers to opposite ends of the track array. While this approach does not completely separate particles---some mixing can occur, particularly when the state size is small---it improves performance as long as the number of tracks is sufficiently large.

Since initializing the geometry state from the track's position can be expensive, another optimization aims to reuse the parent's geometry state by immediately initializing a secondary in the parent's track slot if it is vacant. This provides a substantial performance boost for complex geometries. However, it diminishes the benefits of track partitioning by mixing neutral and charged tracks. Consequently, this optimization is not applied here, leading to higher geometry costs. Future work will explore alternative methods to accelerate geometry state initialization and mitigate this penalty.

\subsection{Reducing load imbalance}

\begin{figure}[htbp]
  \centering%
  \includegraphics[width=0.7\linewidth]{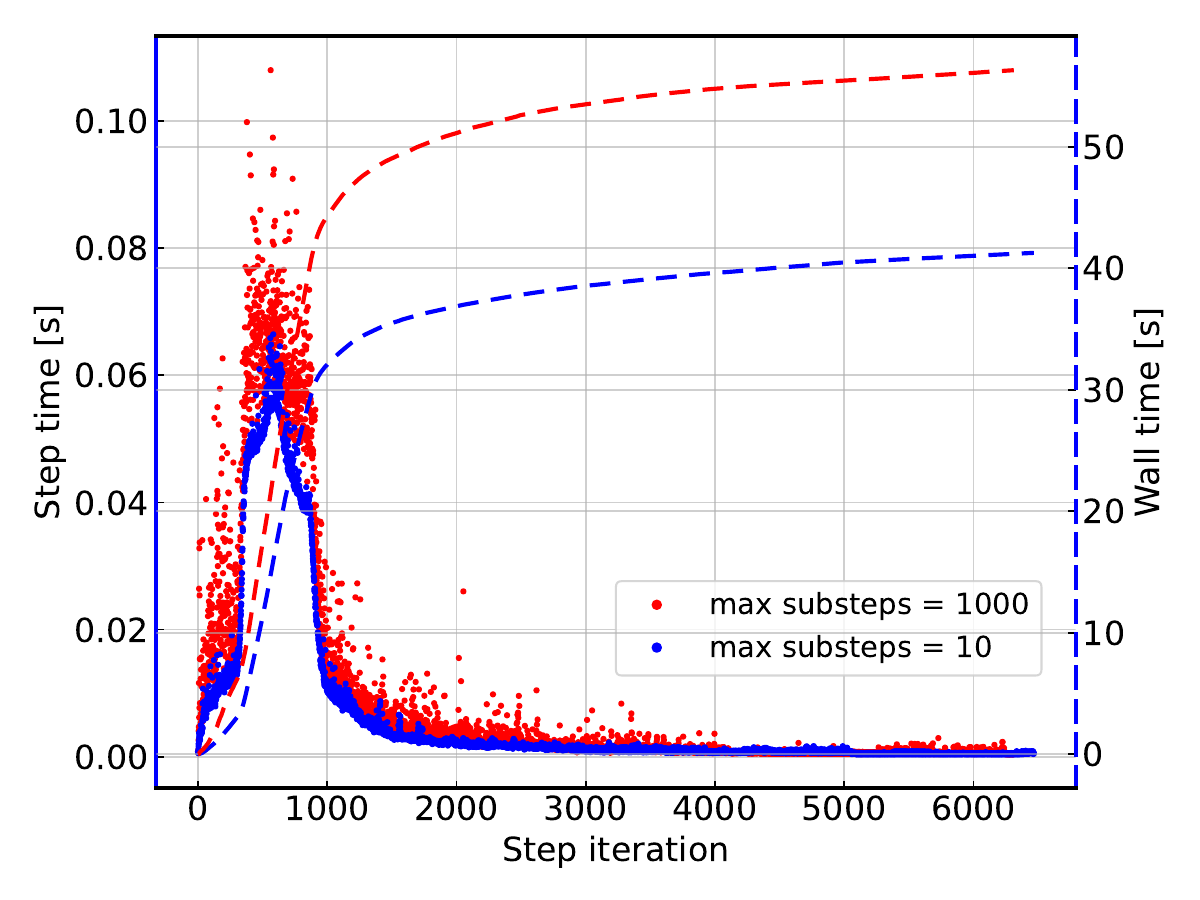}%
  \caption{Simulation of 7 events, each consisting of 1300 10 GeV electron primaries, in the CMS Run 2 geometry under a 3.8 T uniform field. A higher limit on field substeps per step iteration leads to greater load imbalance in the along-step kernel and increased variability in time per step. Reducing the maximum substeps from 1000 to 10 yields a 35\% speedup, despite 98\% of track steps converging in fewer than 10 substeps.}
  \label{fig:field-max-substeps}
\end{figure}

Despite the use of smaller, specialized kernels, load balancing remains a challenge, particularly in the along-step action. The trajectory of a charged particle in a field is approximated by discretizing its curved path into linear substeps, each determined by numerical integration of the equation of motion. While most tracks take only one or two substeps, some ``curling'' tracks may need thousands, leading to significant load imbalance. To mitigate this, a limit is placed on the number of substeps per iteration. When a track reaches this limit, its propagation is paused and resumed in the next iteration. This introduces a tradeoff between kernel execution time and the overall number of step iterations. As shown in Figure~\ref{fig:field-max-substeps}, limiting the substeps to 10 was found to reduce the average along-step kernel time without increasing the total number of iterations beyond statistical fluctuations.

\subsection{Asynchronous memory management}

\Ac{hep} software built on Geant4 parallelizes execution by assigning events to CPU threads. In Celeritas, each CPU thread is mapped to a GPU stream, with all streams executing concurrently on a single \ac{gpu}. For realistic workloads using multiple threads (streams), memory transfer can become a significant bottleneck, particularly when transferring hit data from device to host for processing in user-defined sensitive detectors. While the actual data volume per transfer is small, the frequency is high: hits must be transferred at every step iteration for each event.

By default, synchronous memory operations (e.g., \texttt{[cuda|hip]Memcpy}) enforce a device-wide synchronization, preventing any kernel in any stream from running while the transfer occurs. Given the high transfer frequency, this can severely impact GPU utilization.

To mitigate this issue, we fully migrated all memory operations to the stream-ordered memory API. Thrust kernel calls are issued in the appropriate stream, and a dedicated memory resource ensures stream-ordered memory allocation and transfers. We implemented a \texttt{PinnedMemory} allocator, enabling seamless declaration of pinned \texttt{std::vector} objects for efficient memory transfer, and developed utilities for asynchronous copying between pinned and device memory. These improvements collectively resulted in substantial performance gains, yielding speedups ranging from 10--100\% across our benchmark problems.

\section{Integration with Perfetto}

Perfetto is an open-source tracing tool developed by Google, primarily for the Chromium browser. \emph{Tracing} records detailed, timestamped events throughout a program’s execution, enabling fine-grained performance analysis. In contrast, \emph{profiling} gathers statistical summaries of resource usage by periodically sampling the program. While Perfetto supports both, this work focuses on integrating Celeritas with its tracing \ac{sdk}. Since trace data is too verbose for direct inspection, Perfetto offers analysis tools, including a web-based visualizer and C++/Python libraries for SQL-based analysis.

Perfetto offers two tracing modes, both supported in Celeritas: application tracing and system tracing. In application tracing, only events---slices and counters---from the Perfetto \ac{sdk} are recorded, with the tracing daemon running as a separate thread inside the application. System tracing includes these application events but also captures system-level events via the ftrace kernel API (Linux only). This mode requires administrator privileges and runs the tracing daemon as an external process.

\begin{figure}[htb]
  \centering
  \includegraphics[width=3in]{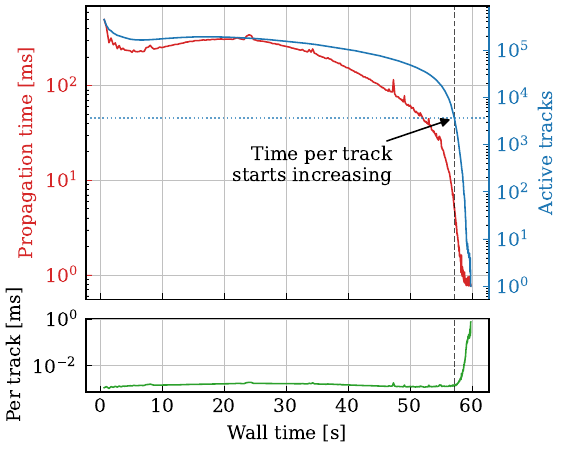}
  \caption{Analysis of Perfetto traces of Celeritas running on an Apple M4 CPU. The plot shows the correlation between propagation time per step iteration (red, along-step kernels) and the number of active tracks. The time per track (green) stays fairly constant, as expected on a CPU, but the overhead is exposed at the tail end of the simulation: when only $\sim 3000$ tracks remain active, the time per track starts to increase.}
  \label{fig:prop-vs-alive}
\end{figure}

Celeritas supports tracing through the CUDA (NVTX)~\cite{nvtx} and AMD (ROCTX)~\cite{roctx} libraries by providing an RAII-wrapper class, \texttt{ScopedProfiling}, around \texttt{[nvtx|roctx]Range[Start|Stop]}. Timing a function requires only instantiating a \texttt{ScopedProfiling} object. To integrate Perfetto, a \texttt{ScopedProfiling} Perfetto back-end is available when neither CUDA nor ROCM are used. Celeritas also supports Perfetto’s counter track events, named numeric values explicitly sampled by the program (e.g., the number of active or queued tracks at each step iteration). Recording counters is straightforward, requiring a single function call with the counter name and value.

To record track events, a Perfetto tracing session must be started. Celeritas provides a RAII-based \texttt{TracingSession} wrapper around Perfetto’s C-style API. Users instantiate a \texttt{TracingSession} object and call its start method when ready to begin tracing; the session automatically finalizes upon destruction. Since Perfetto is bundled with Celeritas as a private dependency, integrating applications do not need to install it separately.

Perfetto traces can be analyzed using its web interface for a high-level overview, such as identifying timing gaps between actions or step iterations. More detailed analysis is possible with the python library and SQL interface, allowing calculations like cumulative time spent per action in each step iteration. For example, Figure \ref{fig:prop-vs-alive} shows how the along-step kernel can be identified as a bottleneck and correlated with the number of active tracks, providing insight into how transport scales with the number of active tracks.

\section{Performance results}

A series of benchmark problems with realistic physics and geometry is used to evaluate the impact of the optimizations applied to Celeritas and analyze the \ac{gpu} thread scaling behavior. Each problem starts with 16 events, each containing 1300 10 GeV electron primaries, chosen to simulate the amount of energy per \acs{lhc} collision deposited in the material. TestEM3 is simulated in the presence of a 1~T uniform magnetic field, and the CMS Run 3 and Run 4 geometries use a 3.8~T uniform field. We use two Celeritas applications to evaluate performance and validate physics: the first integrates with Geant4 for \ac{em} offloading, and the second is a standalone application for independent analysis. Simulations are run using a quarter of a node on the Perlmutter supercomputer~\cite{perlmutter}: a single NVIDIA A100 \ac{gpu} and its 16 associated cores of an AMD EPYC 7763 CPU.

\subsection{\ac{gpu} thread scaling}

\begin{figure}[htbp]
  \centering%
  \includegraphics[width=\linewidth]{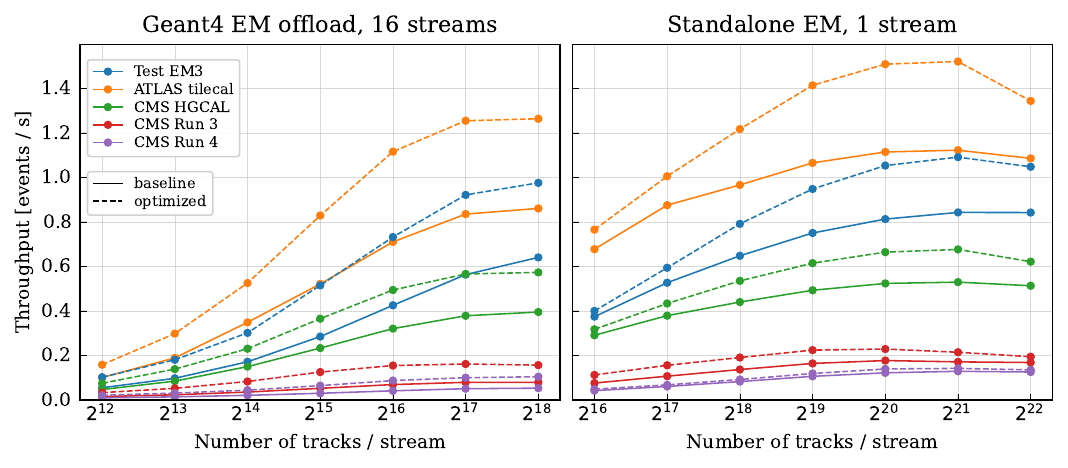}%
  \caption{Throughput scaling with \ac{gpu} thread count on an NVIDIA A100 for the Geant4-integrated and standalone Celeritas applications. The event-parallel model, which runs on multiple \ac{gpu} streams, requires a higher total track count to reach asymptotic throughput. Overall, the applied optimizations yield up to a $2\times$ speedup.}
  \label{fig:gpu-thread-scaling}
\end{figure}

Celeritas integrates with Geant4's event-parallel model by mapping \ac{gpu} streams to CPU threads, with each stream simulating a single event at a time. Achieving asymptotic throughput in this configuration requires a higher total number of tracks, as the number of tracks per stream dictates overall performance. In contrast, when running Celeritas as a standalone application, events are mixed in a single track array and simulated on the default stream. Fewer total tracks are needed to reach the maximum throughput regime, as illustrated in Figure~\ref{fig:gpu-thread-scaling}. Both configurations achieve optimal performance at approximately 2 million \ac{gpu} threads.

\begin{figure}[htbp]
  \centering%
  \includegraphics[width=\linewidth]{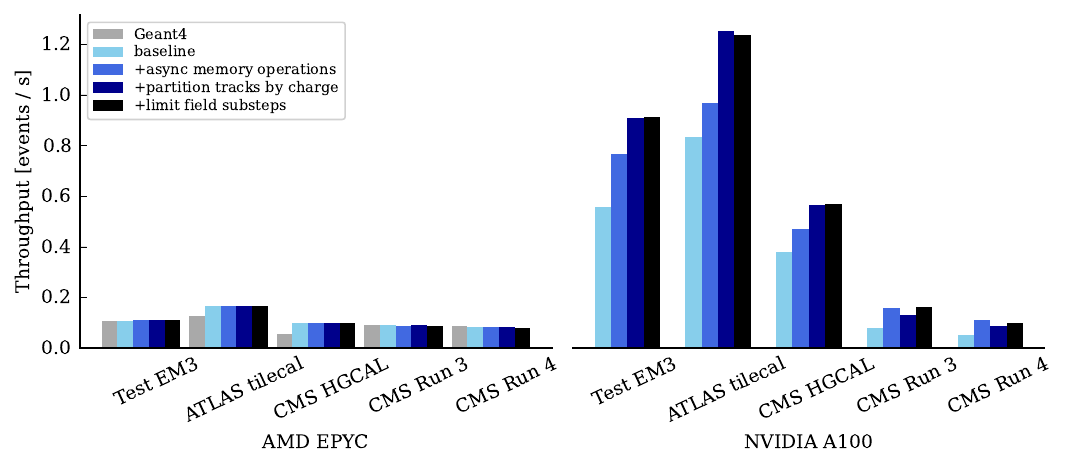}%
  \caption{Throughput for simulating 16 events with the Geant4-integrated application using 16 cores of an AMD EPYC CPU and a single NVIDIA A100 \ac{gpu} with the optimal number of tracks per stream ($2^{12}$ on the CPU and $2^{17}$ on the \ac{gpu}).}
  \label{fig:gpu-threads-g4}
\end{figure}

As expected, increasing geometric complexity reduces throughput due to the higher computational cost of geometry operations. This effect is more pronounced on the \ac{gpu}, as illustrated in Figure~\ref{fig:gpu-threads-g4} and discussed in~\cite{johnson_chep_2023}, where the added complexity disproportionately impacts performance compared to CPU-based transport.

\subsection{Impact of optimizations}

Four primary optimizations were explored to improve the \ac{gpu} performance of Celeritas, with varying degrees of effectiveness depending on problem complexity. The impact of the individual optimizations on the throughput is shown in Figure~\ref{fig:gpu-threads-g4}. As expected, none of these optimizations had any appreciable effect on the CPU performance.

Sorting track state indices and accessing tracks via a lookup array allows for smaller kernel sizes and reduces thread divergence, but it significantly impacts memory coalescing. As a result, this optimization had minimal or even negative effects on performance in most cases. A more successful approach involved partitioning the track data rather than indices by particle type (charged vs. neutral), which resulted in up to a 30\% speedup. However, its impact on problems with greater geometric complexity was limited, as it conflicted with a more effective optimization for geometry initialization in these cases.

Reducing the maximum number of substeps per step iteration in field propagation significantly improved load balancing in the along-step kernel without increasing the total number of step iterations beyond statistical fluctuations. This optimization was particularly effective in the full CMS detector geometries, where it led to up to a 25\% performance improvement.

Migrating all memory operations to the stream-ordered memory API and enabling asynchronous memory transfers between host and device provided the greatest overall performance benefit in an integrated Geant4 configuration. Furthermore, it yielded greater improvements for more complex geometries, with speedups ranging from 15\% for the ATLAS tile calorimeter to $2\times$ for the CMS detector geometries.

\section{Conclusions}

Optimizations to the \ac{gpu}-enabled Celeritas code yield a cumulative $1.5-2\times$ speedup on an NVIDIA A100 across five realistic benchmark problems. Reducing load imbalance in field propagation, partitioning tracks to minimize branch divergence, and eliminating synchronous memory operations lead to maximum speedups of 25\%, 30\%, and 100\%, respectively. Integration of the Perfetto tracing tool enables fine-grained performance analysis, helping identify bottlenecks and concurrency issues. Further performance improvements are expected through optimizations to the geometry routines.

\section*{Acknowledgements}
\begin{spacing}{0.8}
{\small%
This material is based upon work supported by the U.S.~Department of Energy,
Office of Science, Office of Advanced Scientific Computing Research and Office
of High Energy Physics, Scientific Discovery through Advanced Computing (SciDAC)
program.
Work for this paper was supported by Oak Ridge National Laboratory (ORNL), which
is managed and operated by UT--Battelle, LLC, for the U.S. Department of Energy
(DOE) under Contract No.~DE-AC05-00OR22725
and by
Fermi National Accelerator Laboratory, managed and operated by Fermi Forward Discovery Group, LLC under Contract No. 89243024CSC000002
with the U.S. Department of Energy.
This research used resources of the Oak Ridge Leadership Computing Facility at
the Oak Ridge National Laboratory, which is supported by the Office of Science
of the U.S. Department of Energy under Contract No. DE-AC05-00OR22725.
This research used resources of the National Energy Research Scientific
Computing Center (NERSC), a U.S. Department of Energy Office of Science User
Facility located at Lawrence Berkeley National Laboratory, operated under
Contract No. DE-AC02- 05CH11231 using NERSC award HEP-ERCAP-0023868.
}
\end{spacing}

\bibliography{references}

\end{document}